# Five-years Altitude Statistics of Noctilucent Clouds
# Based on Multi-Site Wide-Field Camera Survey

Oleg S. Ugolnikov[a*], Nikolay N. Pertsev[b], Vladimir I. Perminov[b], Ilya S. Yankovsky[c], Dmitry N. Aleshin[d], Ekaterina N. Tipikina[d], Alexander A. Ilyukhin[c], Egor O. Ugolnikov[e], Stanislav A. Korotkiy[f], Olga Yu. Golubeva[g], Andrey M. Tatarnikov[h], Sergey G. Zheltoukhov[h], Alexey V. Popov[i], Alexey M. Sushkov[j], Egor A. Volkov[j], Natalya S. Krapkina[j], Damir I. Yalyshev[c]

*[a] Space Research Institute, Russian Academy of Sciences, Moscow, Russia*
*[b] A.M. Obukhov Institute of Atmospheric Physics, Russian Academy of Sciences, Moscow, Russia*
*[c] Night sky survey project Starvisor.net, Russia*
*[d] K.Tsiolkovsky State Museum of the History of Cosmonautics, Kaluga, Russia*
*[e] Silaeder School, Moscow, Russia*
*[f] "Ka-Dar" observatory, Moscow, Russia*
*[g] Omsk educational institution "GDDYuT", Omsk, Russia*
*[h] Lomonosov Moscow State University, Moscow, Russia*
*[i] Saint-Petersburg State University, Saint-Petersburg, Russia*
*[j] Unaffiliated persons*

*Corresponding author e-mail: ougolnikov@gmail.com

**Abstract**
The results of simultaneous measurements of noctilucent clouds (NLC) position in a number of ground-based locations are presented. Observational data of 14 bright NLC events over 5 years is used for building the altitude maps of cloud fields using triangulation technique updated for multi-location case. Statistical distribution of NLC altitude and its change during the summer season is considered. Mean NLC altitudes are compared with colorimetric technique based on the same data and simple radiation transfer model. This can be used to check the model and estimate the accuracy of single-camera technique of NLC altitude measurements. Results and methods are suggested for net ground-based survey of noctilucent clouds.

## 1. Introduction

The phenomenon of noctilucent clouds was observed for the first time in 1885 (Leslie, 1885; Backhouse, 1885) soon after the strong eruption of Krakatoa in 1883. Direct solar illumination during the deep twilight stage pointed to the unusual position of these clouds, making them the highest above the Earth. In the same year, first attempt of triangulation measurements based on the observations from locations was made: V. Tseraskii and A. Belopolsky used the short base (32 km) to find the altitude of noctilucent clouds: 79 km (Bronsten and Grishin, 1970; Dalin et al., 2012). Jesse (1896) had conducted a number of measurements during the following years to find the mean altitude 82.1 ± 0.1 km. Gadsden and Schröder (1989) had presented the base of triangulation altitude measurements of NLC in 20st century made by a number of authors, the median result was close to 83 km. The same is confirmed by summary of 695 measurements in 1887-1964 referred by Bronsten and Grishin (1970).

Being a rare event in late 19$^{th}$ and early 20$^{th}$ century, noctilucent clouds became brighter and more frequent after that (Thomas and Olivero, 2001). Visual effect was confirmed by increase of total mass of ice in mesosphere measured by SBUV satellite instruments (DeLand and Thomas, 2015), especially in high latitudes. Observed trends were initially related with atmosphere cooling driven by increase of $CO_2$ (Roble and Dickinson, 1989). However, no visual trend of NLC occurrence rate was seen in midlatitudes during the last 60 years (Pertsev et al., 2014; Dalin et al., 2020).

Models (Lübken et al., 2018) show that increase of ice mass and cloud brightness are basically driven by increase of mesospheric $H_2O$, while growing $CO_2$ causes the "atmospheric shrinking" effect when temperature decreases at the fixed geometrical altitude, but remains almost constant at the layer with fixed pressure (Bailey et al., 2021; Mlynczak et al., 2022). It is confirmed by absence of temperature trend in mesopause, above NLC layer (Berger and Lübken, 2011) in midlatitudes during the last decades.

Continuing increase of $CO_2$ in mesosphere in 21$^{st}$ century is the reason of interest for different methods of the observations of atmospheric cooling and shrinking. It can also be contributed by stratospheric ozone depletion (Bremer and Peters, 2008; Walsh and Oliver, 2011). According to model results (Berger and Lübken, 2015; Lübken et al., 2018), the negative trend of clouds altitude about –2 km per century can be observed. This can be fixed by systematical observations during the several dozens of years or by comparison with historical data using the same measurement techniques. However, up to now, efforts of historical data analysis on NLC attitudes did not reveal this trend. Von Zahn (2003) demonstrated a good conservation of the mean NLC altitude during the 20-th century with its variation from 82.2 to 83.6 km, moreover, the large part of this variation seemed to be caused by latitude dependence of NLC altitude. This result is discussed below in this paper.

Triangulation of NLC fields was used for altitude estimation since the beginning of their observations. Using the network of modern photographic devices with fine angular resolution and wide field of view allows increasing the accuracy. Triangulation is used for NLC in the present time (Dubietis et al., 2011; Dalin et al., 2013; Suzuki et al., 2016). Running the procedure for definite fragments of the field, we can build the maps of mean NLC altitude. This method is the most cost-effective and can be the base of spatially expanded system of NLC survey. Multi-band photometric measurements of clouds those are possible with well-calibrated RGB-cameras give the additional opportunity to fix the fields of mean particle size (Ugolnikov, 2024) and estimate the mean altitude by analysis of spectral evolution of the clouds as they are immersed into the shadow of dense atmospheric layers (Ugolnikov, 2023ab).

The primary goal of this work is to present the results of the first years of ground-based camera network and to expand the triangulation procedure to the case of large number of the observation sites. The statistical properties of the altitude distribution of noctilucent clouds are analyzed in comparison with different measurements techniques and historical data.

## 2. Observations

Observations of noctilucent clouds were hold during the summers 2020-2024 by the sample of cameras located in central Russia in Moscow, Tver, Kaluga, and Tula regions (we will call it the central camera set). Basic observational point (55.58°N, 36.56°E, 190 m a.s.l.) is the same as in (Ugolnikov, 2023ab), it is also the southern point of observations of Ugolnikov (2024), where triangulation procedure used here was initially applied to the noctilucent clouds. The cameras in the basic point are hemi-spherical with the field diameter 190° and well-known spectral characteristics of RGB bands (see Ugolnikov, 2023a for description). Angular scale of the image near the zenith is 19 pixels per degree. One more ensemble of NLC was recorded in 2023 by three cameras around Saint-Petersburg (60°N, 30°E), we call it the northern camera set.

Measurements are hold from the evening till the morning during the twilight and the night. Exposure times varied from 0.1 to 8 seconds depending on the camera and twilight stage. Each camera axis direction and optical field parameters are found by star positions astrometry in the nighttime images. High accuracy (mean square residual about 1 pixel or below) is principal for precise cloud position measurements and triangulation. Frame coordinates of several thousands of star positions in different images during the night are measured, it is also principal that the stars must be fixed in all sky areas where NLC images are processed. A new 14-parameter optical field model, the same as in (Ugolnikov, 2024), is used here for all cameras and each night separately. The data with zenith distances up to 70°, where large number of star images can be registered, is being processed. Sky background reduction procedure, initially suggested by Ugolnikov et al. (2021) and improved in (Ugolnikov, 2024), is used.

The choice of observation sites is based on following principles. If the baselength is short, much less than the altitude of the clouds, the error of altitude determination increases. In the opposite case of very long bases the overlapping area of NLC fields fixed by different cameras is reduced. Following this, observational sites surrounded the basic point at the distances of about 50-100 km. Hemispherical cameras were used in several cases. More simple devices with smaller rectangular field of view about 90-120° can be also used if they are installed southwards from the basic point and fix the cloud field over the northern horizon. We note that triangulation procedure uses the correlation analysis and does not require the comparison of absolute brightness, as it is done for particle size estimation or umbral altitude analysis. This allows using the different cameras and different data formats. However, fine angular scale (about 20 degrees per pixel), the same or better than the scale of the camera in the basic observation point, is required.

If noctilucent clouds are expanded covering the most part of the sky hemisphere, RGB data of all-sky cameras basic observation of central set with well-known spectral characteristics can be used to find "umbral" NLC altitudes. The procedure is based on the comparison of observed color indexes of clouds with the data of theoretical radiative transfer model taking into account single scattering of solar emission by cloud particles, refraction, Rayleigh and aerosol extinction, $O_3$ and $NO_2$ absorption, the size of solar disk with darkening to the edge. The procedure is described in (Ugolnikov, 2023a). Having two independent techniques of NLC altitude estimation, we can compare the results based on the same set of observations. This allows checking the quality of the radiative transfer model. Altitude comparison on a single case of bright clouds (Ugolnikov, 2024) had shown the positive difference between umbral and triangulation altitudes, this can be interpreted as the influence of multiple scattering (or the scattering of Earth's limb background by NLC particles). In this paper, we analyze a number of NLC fields to check this assumption.

## 3. NLC triangulation

Triangulation procedure used in this work is principally the same as in (Ugolnikov, 2024) where it was used for observations in two sites with double-camera regime in each location. The cloud position data of these cameras was averaged and then compared in two sites. Here we expand the procedure to the case of unlimited number of sites and cameras. If two or more cameras are installed in the same location, the data is being processed independently, actually increasing the weight of the measurements in this site. Correlations of NLC fields are calculated in pairs, basic (first) camera with each other. All-sky camera with the best quality in the basic observation point is chosen as the first.

Each NLC field registered by one camera of the set in definite location is projected onto the surface at the constant altitude $H_0$ above the sea level using the ellipsoidal model of the Earth. Altitude $H_0$ is considered as *a priori* altitude of noctilucent clouds. It is convenient to use the fixed value for all observations, we choose $H_0$ = 81.33 km following (Ugolnikov, 2024). If NLC fragment is exactly at this altitude, its images will coincide in all fields from any location. Displacements of images are used to find the altitude correction for each fragment.

Fragmentation procedure of NLC field is run for data of the first camera. It is described in details by Ugolnikov (2024) and repeated here without any change. Each fragment surrounds the local maximum of cloud brightness, the strongest maxima where correlation can be found with good accuracy are surrounded by smaller fragments, and finer structure of altitude distribution can be studied there.

Let us consider the coordinate system ($x$, $y$) on the surface with fixed *a priori* altitude of noctiluent clouds $H_0$. The zero point of this system is the projection of the basic observations point to this surface. At the initial stage of the analysis, fields are compared with each other within the two-minute interval, this allows finding the zonal and meridional velocities of cloud patterns. Analysis is similar to (Baumgarten and Fritts, 2014; Ugolnikov, 2024).

Figure 1 shows the distribution of the pattern velocity of the fragments of noctilucent clouds recorded by central camera set. We see that NLC patterns are moving south-westwards, the velocity can reach 100 m/s. While westward motion is typical for summer upper mesosphere, southward velocity component appears in NLC cases, when cold air masses are transported from the polar regions (Fiedler et al., 2011, Gerding et al., 2021).

When the velocity is found, the fields for each camera are added up within the two-minute interval with account of this motion, and the sum is calculated for the middle of this time interval (see Ugolnikov, 2024, for details). These fields for different cameras are compared with each other to find the shift value and cloud fragment altitude. This can be done if the fragment is mostly or totally present in two or more camera frames in different locations, including the first camera in the basic site.

Figure 2 shows the scheme of altitude estimation for the case of two cameras and one NLC fragment. The plumb line from the fragment of noctilucent cloud crosses NLC *a priori* altitude surface in a point with coordinates ($x_N$, $y_N$). The simplified case where $x$-coordinate of both cameras and cloud fragment is equal to zero is painted in the figure.

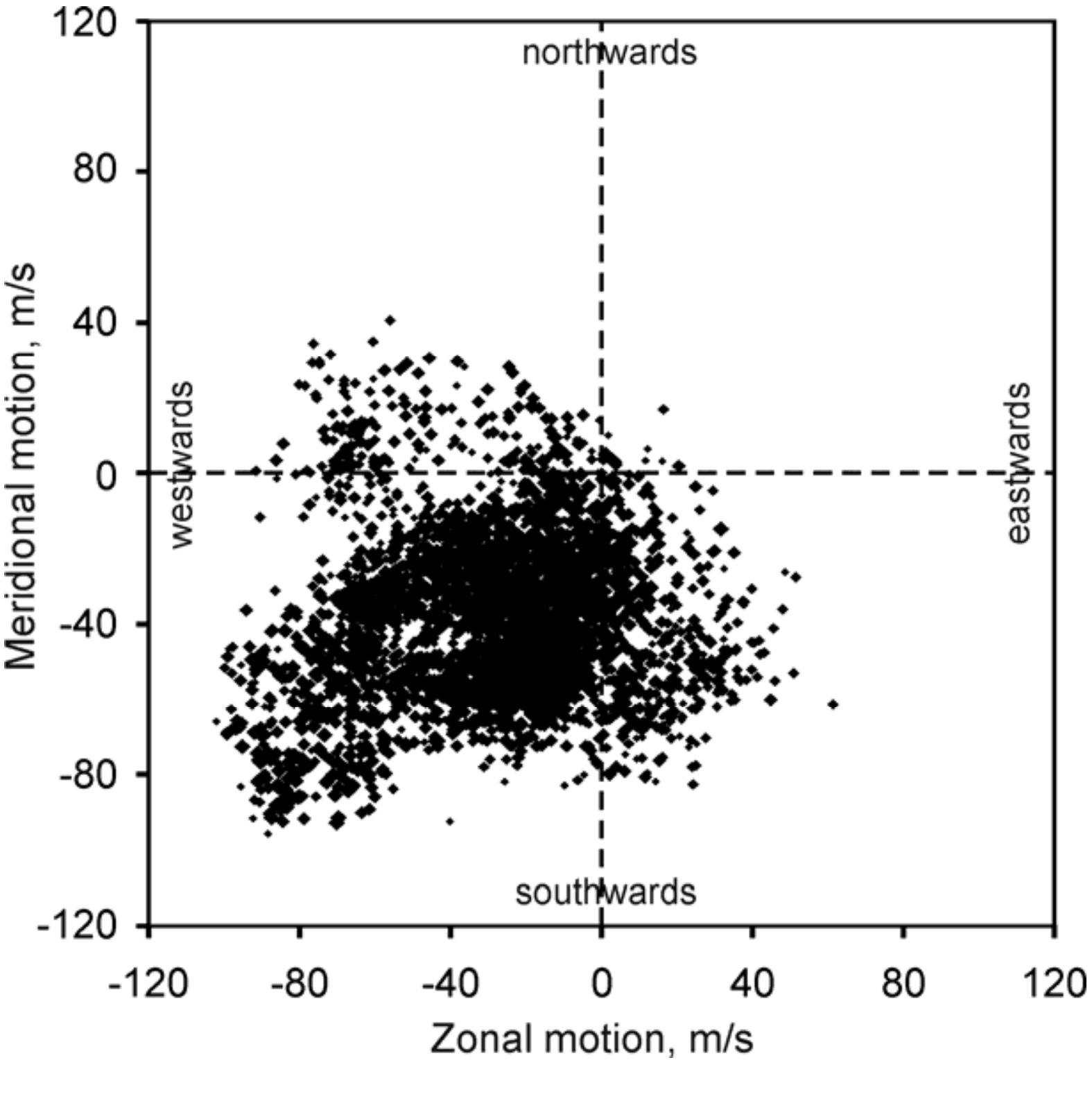

*Figure 1. Pattern motion velocities of the observed fragments of noctilucent clouds. Size of the symbols refers to the brightness of the fragments.*

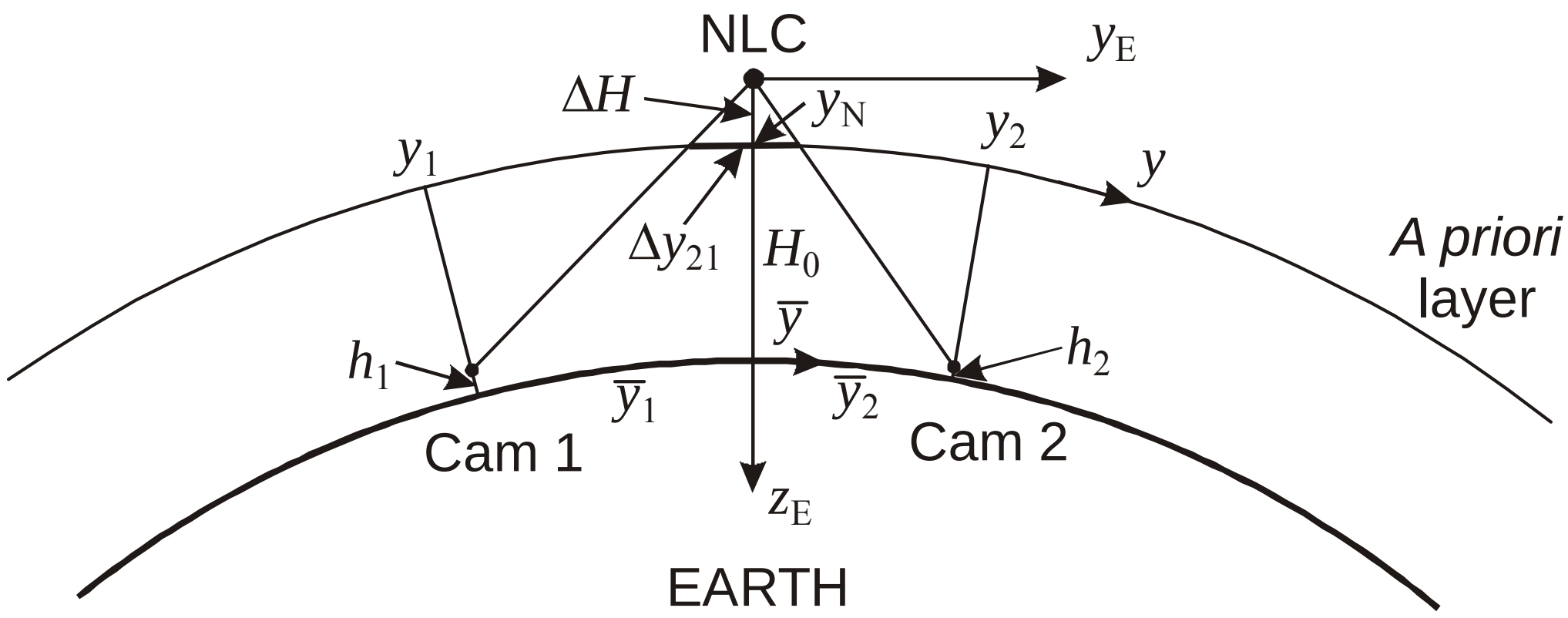

*Figure 2. Graphical scheme of the altitude estimation.*

Assuming the altitude of NLC to be equal to $H_0$ and projecting the picture on *a priori* surface, we run the correlation analysis between camera fields 2 and 1 and find the shift values $\Delta x_{21}$ and $\Delta y_{21}$. In the first assumption, we ignore the curvature of the Earth's surface and small altitudes of observational points $h_i$. For this case, we can write the expressions for the correction of altitude in the first assumption $\Delta H_A$:

$$\Delta H_A\,(x_2 - x_1 - \Delta x_{21}) = H_0 \cdot \Delta x_{21} + C_{AX};$$
$$\Delta H_A\,(y_2 - y_1 - \Delta y_{21}) = H_0 \cdot \Delta y_{21} + C_{AY}. \quad (1)$$

Here $x_{1,2}$ and $y_{1,2}$ are coordinates of camera location projected to the NLC *a priori* layer (we had chosen $x_1 = y_1 = 0$ for the first camera). Rewriting these equations for camera pairs (2-1), (3-1), … ($M$-1), where $M$ is the total cameras number, we find the value of $\Delta H_A$ by the least squares method. Initially unknown constants $C_{AX}$ and $C_{AY}$ are also found to take into account possible errors of NLC fragment position in the frame of the first camera, for which $\Delta x_{11} = \Delta y_{11} = 0$. As it was noticed in (Ugolnikov, 2024), "flat" value of NLC altitude correction is already good, the error rarely exceeds 0.1 km. However, we can clarify the procedure, taking into account the curvature of the Earth and non-zero altitudes of observation sites. A number of coordinate transformations is needed for this.

At this stage we assume the Earth to be spherical with radius $R$. The cloud element is projected to the *a priori* layer at coordinates ($x_N$, $y_N$). We define the layer coordinate system on the Earth's surface analogous to ($x$, $y$) system at *a priori* altitude of NLC. The centre of this system is located on the Earth's surface below the cloud element:

$$\bar{x}_{1,2} = (x_{1,2} - x_N)\frac{R}{R + H_0};$$
$$\bar{y}_{1,2} = (y_{1,2} - y_N)\frac{R}{R + H_0}. \quad (2)$$

Then we consider the Cartesian coordinate system ($x_E$, $y_E$, $z_E$) with center at the approximated NLC fragment position with altitude $H_0 + \Delta H_A$ above the Earth surface. Axis $z_E$ is directed downwards to the Earth's center, and $y_E$ is parallel to the Earth's meridian below the cloud (see Figure 2). The coordinates of the observation points are:

$$x_{E1,2} = (R + h_{1,2})\sin(\bar{x}_{1,2}/R) \approx \bar{x}_{1,2}\cdot(1 + h_{1,2}/R);$$
$$y_{E1,2} = (R + h_{1,2})\sin(\bar{y}_{1,2}/R) \approx \bar{y}_{1,2}\cdot(1 + h_{1,2}/R);$$
$$z_{E1,2} = R + H + \Delta H_A - (R + h_{1,2})\cos\left(\sqrt{\bar{x}_{1,2}^2 + \bar{y}_{1,2}^2}/R\right) \approx H + \Delta H_A + \frac{\bar{x}_{1,2}^2 + \bar{y}_{1,2}^2}{2R} - h_{1,2}. \quad (3)$$

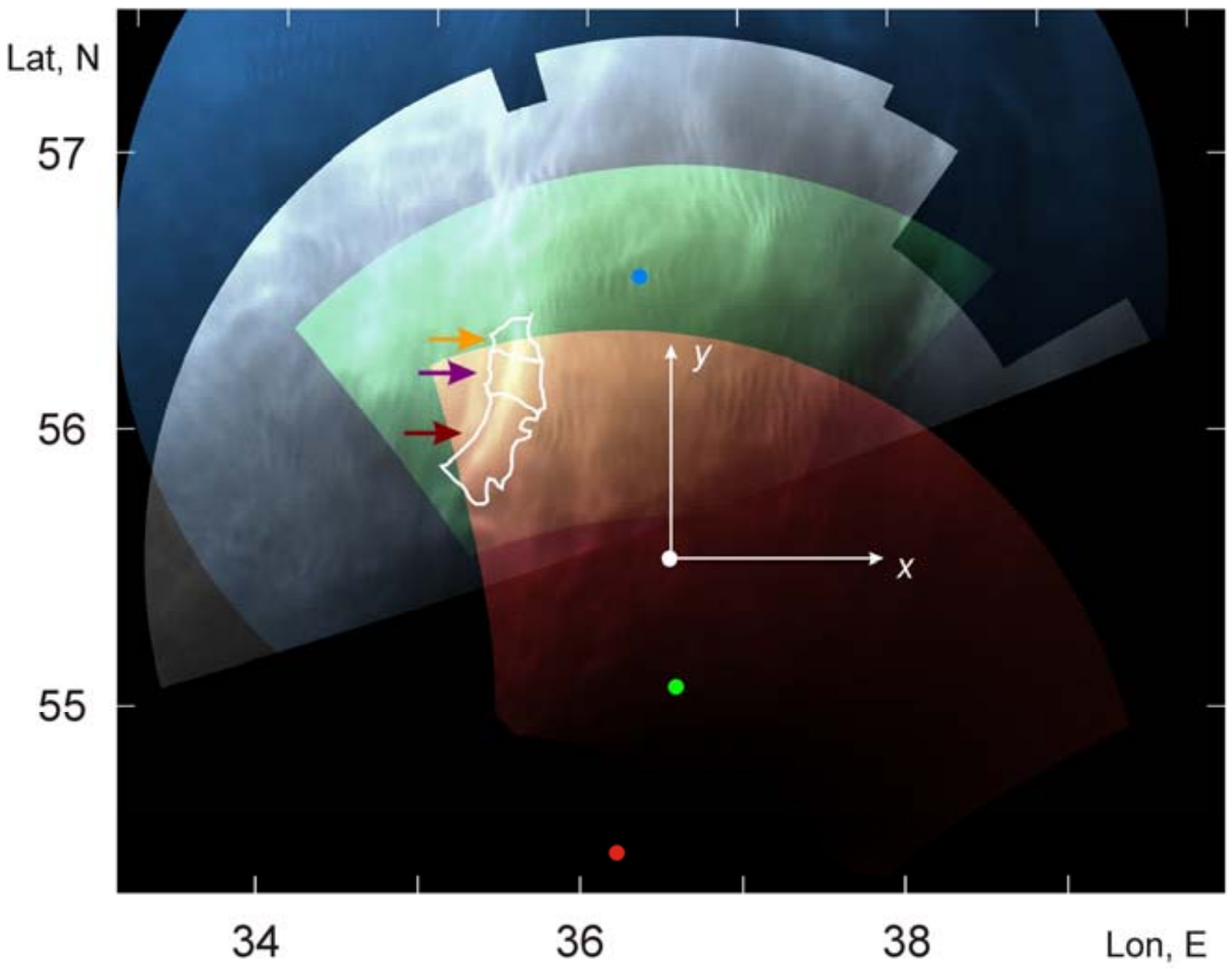

*Figure 3. NLC field in the evening, July 3, 2023, 19:48 UT, projected onto a priori layer from four observation locations signed with the same color. Three bright NLC fragments being analyzed in the next figure are shown as the example.*

Approximations in the last equation can be used for camera altitudes below 1 km and clouds observed high above the horizon. However, exact analysis is necessary if mountain or balloon-borne experiment is planned, and NLC near the horizon are considered. The shift of the image 2 relatively image 1 is small, and we can consider it as a straight line. Projections of this displacement on the axes $x$ and $y$ at *a priori* layer are:

$$\Delta x_{21} = \Delta H \cdot \left( \frac{x_{E2}}{z_{E2}} - \frac{x_{E1}}{z_{E1}} \right) + C_X;$$

$$\Delta y_{21} = \Delta H \cdot \left( \frac{y_{E2}}{z_{E2}} - \frac{y_{E1}}{z_{E1}} \right) + C_Y. \quad (4)$$

Again, terms $C_X$ and $C_Y$ are added to fix possible position errors of the fragment in the first camera. The value of $\Delta H$ is found by the data of the camera set by the least squares method. Then it can be used as approximate value $\Delta H_A$ to repeat the procedure, 3 or 4 iterations are enough for this process.

Figure 3 shows the example of NLC field projections onto *a priori* layer basing on the observations from four locations. The fields are painted in different colors for clarity, positions of observational sites are shown by the dots with the same colors. These fields visually coincide, but actually correlation analysis has revealed the displacements of order of 1 km. Since the observation sites during this twilight were practically aligned from north to south, the fields are shifted along the $y$-axis. However, $x$-axis analysis was also conducted, contributing to the errors of altitude definition.

Dependencies of $y$-shift relatively the basic camera (white field) on the site position along the $y$-axis are shown in Figure 4 for three bright NLC fragments contoured in Figure 3. Dashed lines correspond to the second expression in equation (1), which is used to find the altitude correction $\Delta H_A$ and then $\Delta H$, the difference between them is less than 0.02 km for these three fragments. Running this procedure for various fragments and moments of time, we build the maps of mean NLC altitude during the twilight.

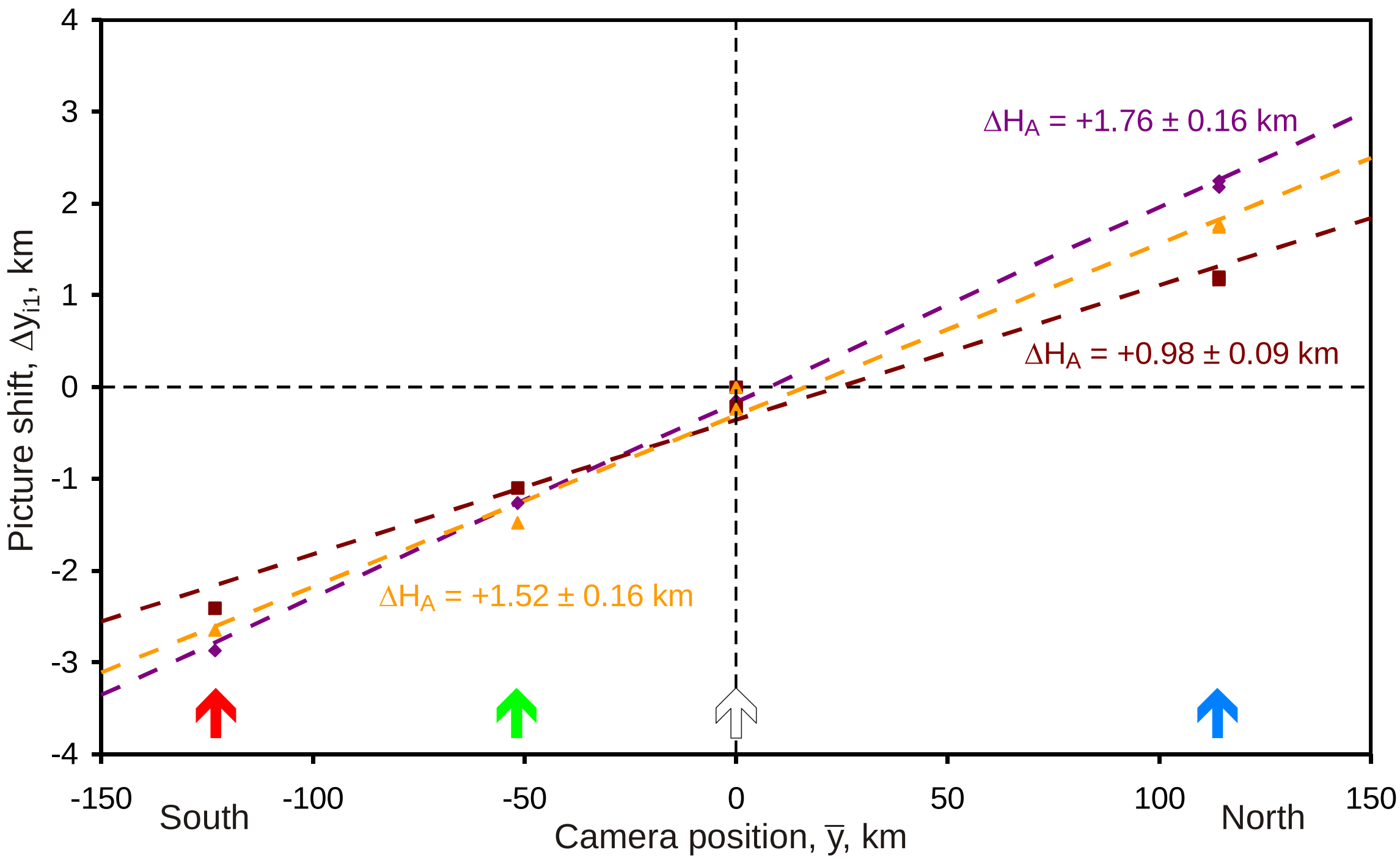

*Figure 4. Displacement of the NLC field on a priori surface compared the first camera in basic observational point for three fragments shown in Figure 3 (the same color). Arrows show the positions of observation sites, the same color as in Figure 3. Note that double-camera measurements were conducted in "white" and "blue" locations, the dots corresponding to the common site are separate but close to each other.*

## 4. NLC altitude statistics

Figure 5 shows the maps of NLC altitudes built by triangulation technique, one map for each observed twilight in 2020-2024. Positions and character angles of view of the cameras involved to the measurements during each twilight are shown, all-sky cameras are marked by the circles.

We see that NLC altitudes are almost totally in the interval from 79 to 84 km. Statistical distribution of NLC fragments is presented in Figure 6 in square-weighted and brightness-weighted forms: distributions of sums of square and sums of brightness of clouds fragments with definite altitudes. It is expected that brightness-weighted distribution is shifted to the lower altitudes, since NLC ice particles are characterized by sharp dependence of scattering coefficient on particle radius. The largest particles form in the bottom of the layer where the temperature is below the frost point (Fiedler et al., 2003; Baumgarten et al., 2010). Altitude-size relation for NLC particles were confirmed by limb satellite measurements (von Savigny et al., 2005). Altitude difference between mesopause and the layer of maximum ice density reaches 4 km during the summer temperature minimum epoch, as it was registered by satellite observations (Li et al., 2024).

Lower slope of brightness-weighted altitude profile is sharper than upper slope, the same is predicted by models (Turco et al., 1982; Rapp and Thomas, 2006; Lübken et al., 2007) and seen in lidar (von Zahn et al., 2004; Ridder et al., 2017; Schäfer et al., 2020) and satellite results (Christensen et al., 2016). However, the mean profile obtained by NLC triangulation is about 1-1.5 km lower than lidar and satellite profiles. It is also below the typical profiles of polar mesosphere summer echoes (Li et al., 2010).

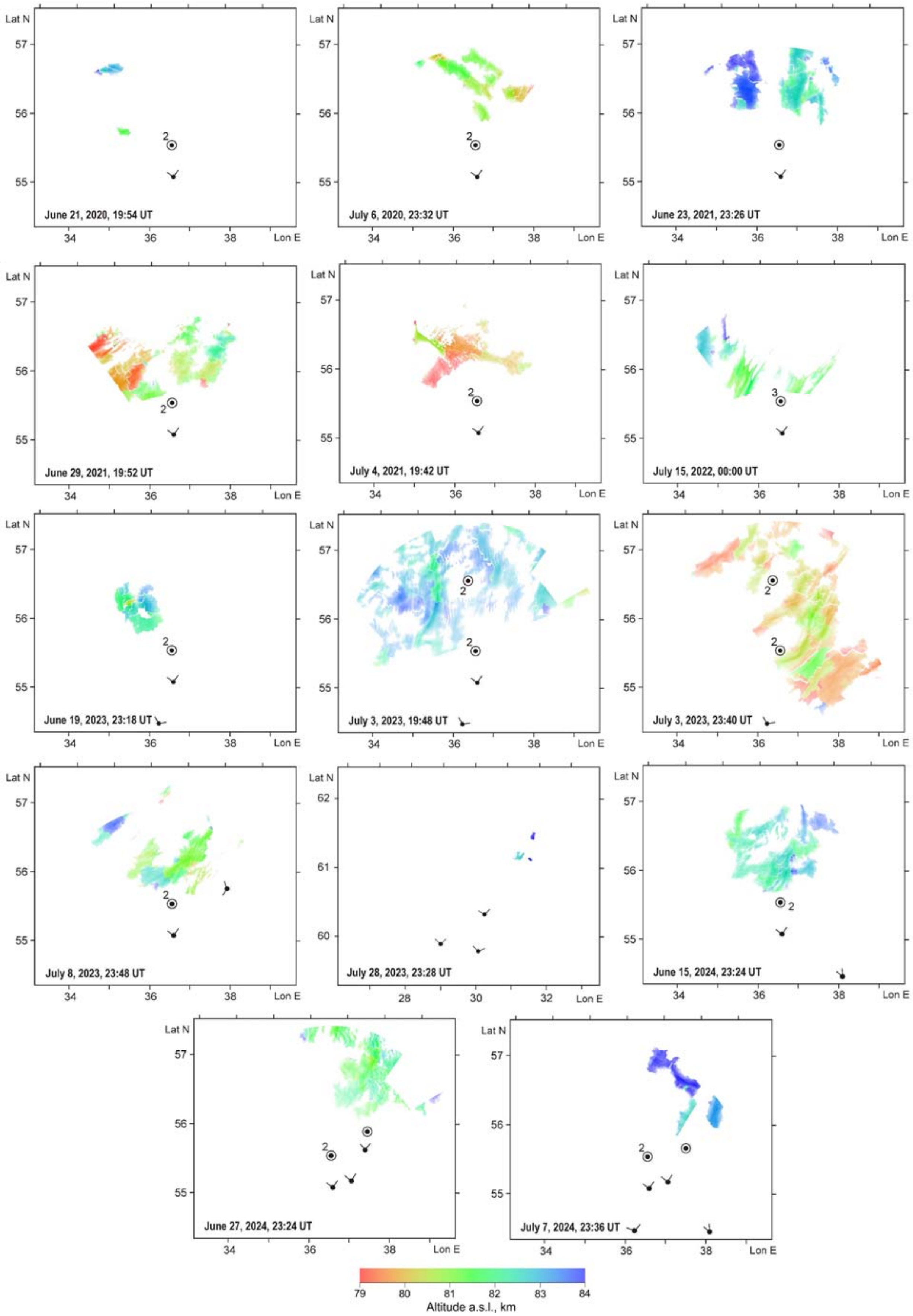

*Figure 5. Altitude maps of NLC for different observation dates. Camera positions and fields of view are shown. Digits mean the number of cameras in the same site if it is more than one.*

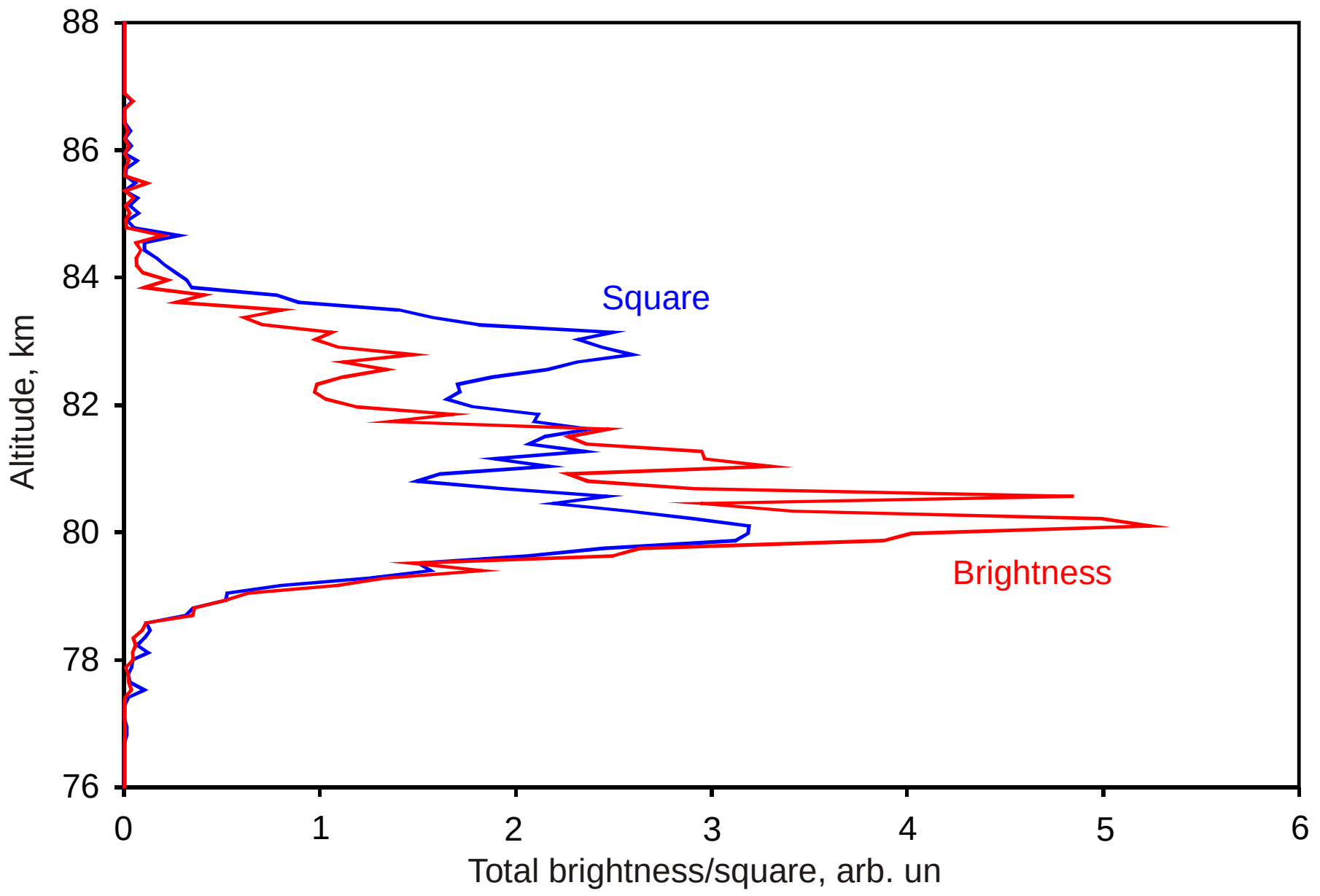

*Figure 6. Brightness- and square-weighted distribution of measured NLC altitudes in 2020-2024.*

Altitudes of noctilucent clouds are further averaged over the space and time for each twilight. These mean altitudes change during the summer cold mesosphere season, as we can see in Figure 7. Close to the moment of temperature minimum, from last days of June to early July, altitudes vary over the whole interval from 79 to 84 km. During the early season stage in the middle of June, the clouds appear only above 82 km according to triangulation, since the frost layer rarely reaches lower altitudes. This behavior of mean NLC altitude with minimum at the mid-season was noticed in limb satellite measurements (Li et al., 2024) and reflects the mean temperature distribution, as it was noticed in lidar results and model calculations (Lübken et al., 2008).

If NLC field observed by all-sky RGB camera in basic observation point is expanded covering the significant part of the sky, we have the alternate opportunity to estimate the mean altitude using the color measurements of clouds as they immerse into the shadow of ozone layer and then troposphere ("umbral" altitude, Ugolnikov, 2023a) and comparison with radiative scattering model. Results are plotted in Figure 8, the agreement is seen. Umbral altitudes are also shown in Figure 7 for the twilights when they were measured. The comparison can be the test of contribution of multiple scattering in observed intensity of NLC. This effect can lead to overestimation of umbral altitude and upward shift of experimental data in the Figure 8. It was noticed during the night of July 3, 2023 (Ugolnikov, 2024), but here we see that the umbral altitude does not significantly exceed the triangulation altitude, the mean difference with account of errors is about 0.5 km. This can be used as the numerical criterion for radiative transfer models for NLC conditions and confirms the possibility to use two independent methods to measure the NLC altitude by ground-based camera network.

## 5. Discussion and conclusion

In this paper, we have introduced the network of photographic and photometric cameras at the distance 50-100 km from each other. This is optimal for triangulation survey of noctilucent clouds. Altitude measurements of NLC are actual, since negative trend related with atmosphere shrinking is possible. As found by models (Lübken et al., 2018), it was basically driven by increase of $CO_2$ in mesosphere.

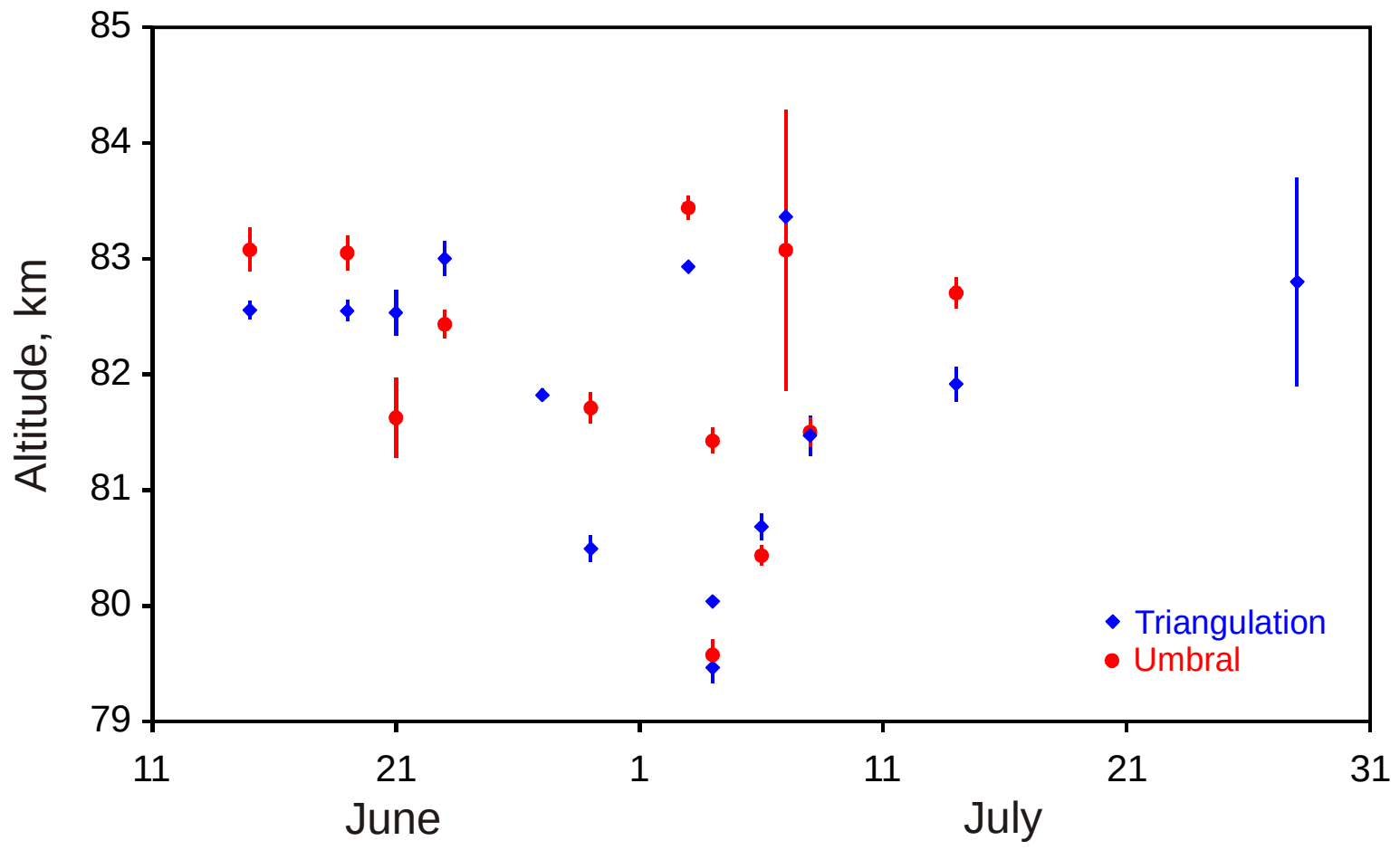

*Figure 7. Mean NLC altitude depending on the observation date in 2020-2024.*

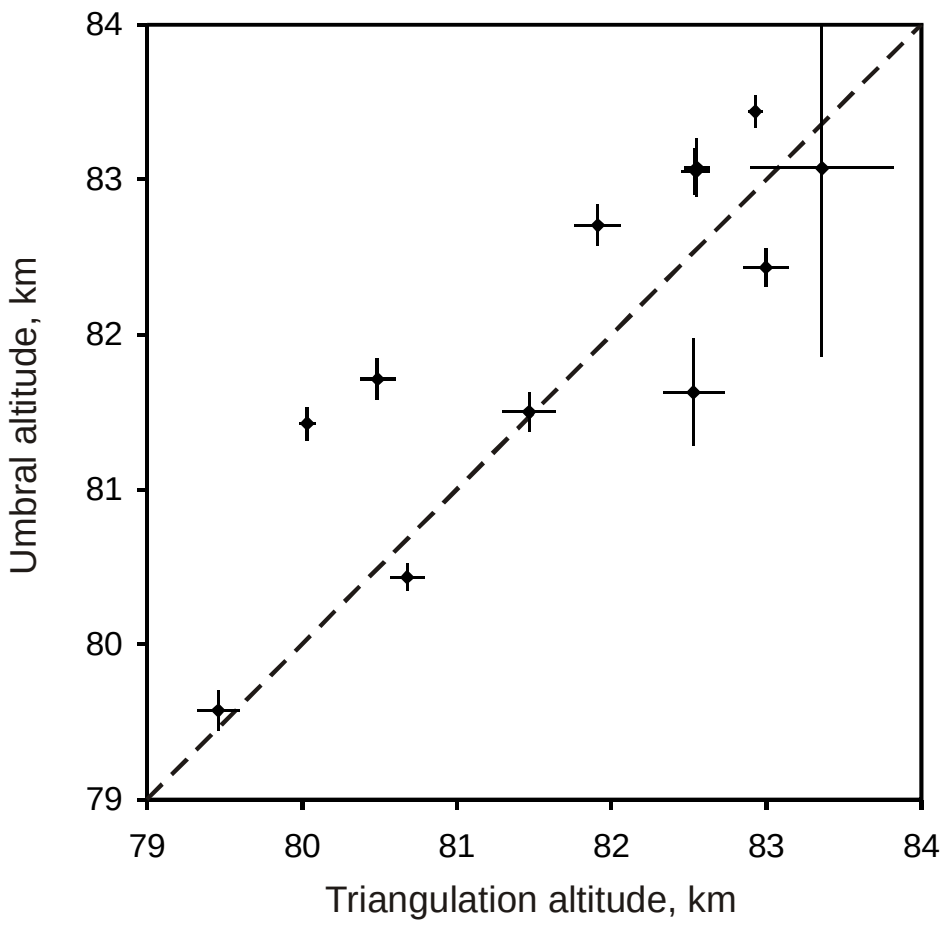

*Figure 8. Comparison of mean triangulation and umbral altitudes of NLC by the data of observations in 2020-2024.*

Possible problem in long-term trend study of NLC is related with change of measuring techniques and difference of altitudes depending on the method. It is known that bright visible clouds are usually lower than mean altitude measured by lidars (Fiedler et al., 2005), the difference can reach 2 km (Gerding et al., 2021). This is related with higher detection threshold for visible NLC (Baumgarten et al., 2009) and sharper dependence of scattering at small angles on the particle size compared with backscattering fixed by lidars. Visible NLC are principally contributed by large particles in the bottom of the frost layer in the mesosphere. This effect can be a possible reason of stable NLC altitude as the result of von Zahn (2003), where historical triangulation data was compared with lidar measurements near the end of 20st century. We also add that mean NLC altitude in mid-latitudes is typically lower than in high latitudes where polar mesospheric clouds are mostly studied (Lübken et al., 2008).

Square-weighted average altitude of NLC is found to be 81.4 km, brightness-weighted average (or mean optical) altitude is 80.9 km. The height of brightest clouds is 80.1 km. These values are below the historical triangulation values by Jesse (1896) and visual altitudes estimated during 20st century (Gadsden and Schröder, 1989). Further observations are needed for confirming this as the display of negative trend of altitude of NLC. We have to take into account the significant season dependence of NLC altitude, here all clouds registered before or during the solstice were higher than 82 km.

Another goal of the paper was to check the “umbral” technique of mean altitude estimation suggested by Ugolnikov (2023a). It is based on the same photometric data of 2020-2024 fixed by all-sky cameras in basic site of the central camera set, but the principle of altitude measurement is completely different. Comparison of the results shows the good agreement, average excess of umbral altitude possibly related with multiple scattering contribution is about 0.5 km. Using this result, we can update the radiative transfer model and improve the accuracy of alternate estimation of NLC by ground camera survey.

RGB-camera network can be used not only for altitude measurements, but also for building the maps of mean particle size (Ugolnikov, 2024). This requires accurate camera calibration, using a number of the same cameras is optimal. Now we have such data only for the night of July 3, 2023, described in the paper referred above, but further updates are planned. In the case of bright and well-structured clouds, we can also update the fragmentation procedure and analyze the tiny cloud fragments, fixing their horizontal and vertical motion and evolution of microphysical properties.

This will be a good test for the models of ice nucleation, evolution and evaporation in the mesosphere.

**Acknowledgments**

Authors would like to thank Natalia A. Abakumova (K.Tsiolkovsky State Museum of the History of Cosmonautics, Kaluga), Nadezhda N. Shakvorostova and Sergey V. Pilipenko (Astro-Space Center of Lebedev's Physical Institute, Moscow), Mikhail V. Kuznetsov (Sternberg Astronomical Institute of Moscow State University), Evgeny G. Merzlyakov, Sergey A. Kulikov, Vyacheslav B. Ignatiev, Kirill O. Chepurnoy, and Kirill O. Ugolnikov for their help in the measurement organization and process. Oleg S. Ugolnikov is supported by Russian Science Foundation, grant 23-12-00207.

**References**

Backhouse, T. W., 1885. The luminous cirrus cloud of June and July. Meteorological Magazine, 20, 133.

Bailey, S.M., Thurairajah, B., Hervig, M.E., Siskind, D.E., Russell, J.M. III, Gordley, L.L., 2021. Trends in the polar summer mesosphere temperature and pressure altitude from satellite observations. J. Atm. Sol-Terr. Phys. 220, 105650. https://doi.org/10.1016/j.jastp.2021.105650.

Baumgarten, G., et al., 2009. The noctilucent cloud (NLC) display during the ECOMA/MASS sounding rocket flights on 3 August 2007: morphology on global to local scales. Ann. Geophys. 27, 953-965. www.ann-geophys.net/27/953/2009/.

Baumgarten, G., Fiedler, J., Rapp, M. 2010. On microphysical processes of noctilucent clouds (NLC): observations and modeling of mean and width of the particle size-distribution. Atm. Chem. Phys. 10, 6661-6668. https://doi.org/10.5194/acp-10-6661-2010.

Baumgarten, G., Fritts, D.C., 2014. Quantifying Kelvin-Helmholtz instability dynamics observed in noctilucent clouds: 1. Methods and observations. J. Geophys. Res. Atmos. 119, 9324–9337. https://doi.org/10.1002/2014JD021832.

Berger, U., Lübken, F.-J., 2011. Mesospheric temperature trends at mid-latitudes in summer. Geophys. Res. Lett. 38, L22804. https://doi.org/10.1029/2011GL049528.

Berger, U., Lübken, F.-J., 2015. Trends in mesospheric ice layers in the Northern Hemisphere during 1961-2013. J. Geophys. Res. Atmos. 120, 11277-11298. https://doi.org/10.1002/2015JD023355.

Bremer, J., Peters, D., 2008. Influence of stratospheric ozone changes on long-term trends in the meso- and lower thermosphere. J. Atm. Sol-Terr. Phys. 70, 1473-1481. https://doi.org/10.1016/j.jastp.2008.03.024.

Bronshten, V.A., Grishin, N.I., 1970. Noctilucent clouds. Nauka, Moscow (in Russian).

Christensen, O.M., Benze S., Eriksson, P. Gumbel, J., Megner, L., Murtagh, D.P., 2016. The relationship between polar mesospheric clouds and their background atmosphere as observed by Odin-SMR and Odin-OSIRIS. Atm. Chem. Phys. 16, 12587-12600. https://doi.org/10.5194/acp-16-12587-2016.

Dalin, P., Pertsev, N., Romejko, V., 2012. Notes on historical aspects on the earliest known observations of noctilucent clouds. Hist. Geo Space Sci. 3, 87-97. https://doi.org/10.5194/hgss-3-87-2012.

Dalin, P., Connors, M., Schofield, I., Dubietis, A., Pertsev, N., Perminov, V., Zalcik, M., Zadorozhny, A., McEwan, T., McEachran, I., Grønne, J., Hansen, O., Andersen, H., Frandsen, S., Melnikov, D., Romejko, V., Grigoryeva, I., 2013. First common volume ground-based and space measurements of the mesospheric front in noctilucent clouds. Geophys. Res. Lett. 40, 6399-6404. https://doi.org/10.1002/2013GL058553.

Dalin, P., Perminov, V., Pertsev, N., Romejko, V., 2020. Updated long-term trends in mesopause temperature, airglow emissions, and noctilucent clouds. J. Geophys. Res. Atmos. 125, e2019JD030814. https://doi.org/10.1029/2019JD030814.

DeLand, M.T., Thomas, G.E., 2015. Updated PMC trends derived from SBUV data. J. Geophys. Res. 120, 2140-2166. https://doi.org/10.1002/2014JD022253.

Dubietis, A., Dalin, P., Balciunas, R., Černis, K., Pertsev, N., Sukhodoev, V., Perminov, V., Zalcik, M., Zadorozhny, A., Connors, M., Schofield, I., McEwan, T., McEachran, I., Frandsen, S., Hansen, O., Andersen, H., Grønne, J., Melnikov, D., Manevich, A., Romejko, V., 2011. Noctilucent clouds: modern ground-based photographic observations by a digital camera network. Appl Opt. 50, F72-F79. https://doi.org/10.1364/AO.50.000F72.

Fiedler, J., Baumgarten, G., von Cossart, F., 2003. Noctilucent clouds above ALOMAR between 1997 and 2001: Occurrence and properties. J. Geophys. Res. 108, D8, 8453, https://doi.org/:10.1029/2002JD002419.

Fiedler, J., Baumgarten, G., von Cossart, F., 2005. Mean diurnal variations of noctilucent clouds during 7 years of lidar observations at ALOMAR. Ann. Geophys. 23, 1175-1181. https://angeo.copernicus.org/articles/23/1175/2005/angeo-23-1175-2005.pdf.

Fiedler, J., Baumgarten, G., Berger, U., Hoffmann, P., Kaifler, N., Lübken, F.-J., 2011. NLC and the background atmosphere above ALOMAR. Atm. Chem. Phys. 11, 5701-5717. http://doi.org/10.5194/acp-11-5701-2011.

Gadsden, M., Schröder, W., 1989. Nocilucent Clouds. Springer-Verlag, Berlin.

Gerding, M., Baumgarten, G., Zecha, M., Lübken, F.-J., Baumgarten, K., Latteck, R., 2021. On the unusually bright and frequent noctilucent clouds in summer 2019 above Northern Germany. J. Atmos. Sol. Terr. Phys. 217, 105577. https://doi.org/10.1016/j.jastp.2021.105577.

Jesse O., 1896. Die Höhe der leuchtenden Nachtwolken. Astronomische Nachrichten. 40, 161-168.

Leslie, R.C., 1885. Sky glows. Nature. 32, 245.

Li, Q., Rapp, M., Rottger, J., Latteck, R., Zecha, M., Strelnikova, I., Baumgarten, G., Hervig, M., Hall, C., and Tsutsumi, M., 2010. Microphysical parameters of mesospheric ice clouds derived from calibrated observations of polar mesosphere summer echoes at Bragg wavelengths of 2.8 m and 30 cm. J. Geophys. Res. 115, D00I13. https://doi.org/10.1029/2009JD012271, 2010.

Li, Y., Gao, H., Sun, Sh., Li, X., 2024. Correlation between peak height of polar mesospheric clouds and mesopause temperature. Atmosphere. 15, 1149. https://doi.org/10.3390/atmos15101149.

Lübken, F.-J., Rapp, M., Strelnikova, I., 2007. The sensitivity of mesospheric ice layers to atmospheric background temperatures and water vapor. Adv. Space Res. 40, 794-801. https://doi.org/10.1016/j.asr.2007.01.014.

Lübken, F.-J., Baumgarten, G., Fiedler, J., Gerding, M., Höffner, J., Berger, U., 2008. Seasonal and latitudinal variation of noctilucent cloud altitudes. Geophys. Res. Lett. 35, L06801, http://dx.doi.org/10.1029/2007GL032281.

Lübken, F.-J., Berger, U., Baumgarten, G., 2018. On the anthropogenic impact on long-term evolution of noctilucent clouds. On the anthropogenic impact on long-term evolution of noctilucent clouds. Geophys. Res. Lett. 45, 6681-6689. https://doi.org/10.1029/2018GL077719.

Mlynczak, M.G., Hunt, L.A., Garcia, R.R., Harvey, V.L., Marshall, B.T., Yue, J., Mertens, C.J., Russell, J.M. III., 2022. Cooling and contraction of the mesosphere and lower thermosphere from 2002 to 2021. J. Geophys. Res. Atmos. 127, D036767. https://doi.org/10.1029/2022JD036767.

Pertsev, N., Dalin, P., Perminov, V., Romejko, V., Dubietis, A., Balčiunas, R., Černis, K., Zalcik, M., 2014. Noctilucent clouds observed from the ground: sensitivity to mesospheric parameters and long-term time series. Earth Plan. Space. 66, 98. https://doi.org/10.1186/1880-5981-66-98.

Rapp, M., Thomas, G.E., 2006. Modeling the microphysics of mesospheric ice particles: assessment of current capabilities and basic sensitivities. J. Atmos. Solar Terr. Phys. 68, 715–744.

Ridder, C., Baumgarten, G., Fiedler, J., Lübken, F-J., Stober, G., 2017. Analysis of small-scale structures in lidar observations of noctilucent clouds using a pattern recognition method. J. Atmos. Sol. Terr. Phys. 162, 48-56. http://dx.doi.org/10.1016/j.jastp.2017.04.005.

Roble, R.G., Dickinson, R.E., 1989. How will changes in carbon dioxide and methane modify the mean structure of the mesosphere and thermosphere? Geophys. Res. Lett. 16, 1441-1444. https://doi.org/10.1029/GL016i012p01441.

Schäfer, B., Baumgarten, G., Fiedler, J., 2020. Small-scale structures in noctilucent clouds observed by lidar. J. Atm. Sol. Terr. Phys. 208, 105384. https://doi.org/10.1016/j.jastp.2020.105384.

Suzuki, H., Sakanoi, K., Nishitani, N., Ogawa, T., Ejiri, M.K., Kubota, M., Kinoshita, T., Murayama, Y., Fujiyoshi, Y., 2016. First imaging and identification of a noctilucent cloud from multiple sites in Hokkaido (43.2–44.4°N), Japan. Earth Plan. Space. 68, 182. https://doi.org/10.1186/s40623-016-0562-6.

Thomas, G.E., Olivero J., 2001. Noctilucent clouds as the possible indicators of global change in the mesosphere. Advances in Space Research, 28, 937-946.

Turco, R.P., Toon, O.B., Whitten, R.C., Keesee, R.G., Hollenbach, D., 1982. Noctilucent clouds: simulation studies of their genesis, properties and global influences. Plan. Space Sci. 30, 1147-1181.

Ugolnikov, O.S., Kozelov, B.V., Pilgaev, S.V., Roldugin, A.V., 2021. Retrieval of particle size distribution of polar stratospheric clouds based on wide-angle color and polarization analysis. Plan. Space Sci. 200, 105213. https://doi.org/10.1016/j.pss.2021.105213.

Ugolnikov, O.S., 2023a. Altitude and particle size measurements of noctilucent clouds by RGB photometry: Radiative transfer and correlation analysis. J. Quant. Spec. Rad. Trans. 296, 108433. https://doi.org/10.1016/j.jqsrt.2022.108433.

Ugolnikov, O.S., 2023b. Cross-wave profiles of altitude and particle size of noctilucent clouds in the case of one-dimensional small-scale gravity wave pattern. J. Atm. Sol. Terr. Phys. 243, 106024. https://doi.org/10.1016/j.jastp.2023.106024.

Ugolnikov, O.S., 2024. Noctilucent clouds altitude and particle size mapping based on spread observations by ground-based all-sky cameras. J. Atm. Sol. Terr. Phys. 259, 106242. https://doi.org/10.1016/j.jastp.2024.106242.

Walsh, P.L., Oliver, W.L., 2011. Is thermospheric long-term cooling due to $CO_2$ or $O_3$? Ann. Geophys. 29, 1779–1782. https://www.ann-geophys.net/29/1779/2011/.

Von Savigny, C., Petelina, S.V., Karlsson, B., Llewellyn, E.J., Degenstein, D.A., Lloyd, N.D., Burrows, J.P., 2005. Vertical variation of NLC particle sizes retrieved from Odin/OSIRIS limb scattering observations. Geophys. Res. Lett. 32, L07806. https://doi.org/10.1029/2004GL021982.

Von Zahn, U. , 2003. Are noctilucent clouds truly a «miner's canary» for global change? EOS. Trans. AGU. 84, 261-264. https://doi.org/10.1029/2003EO280001.

Von Zahn, U., Baumgarten, G., Berger, U., Fiedler, J., Hartogh, P., 2004. Noctilucent clouds and the mesospheric water vapour: the past decade. Atm. Chem. Phys. 4, 2449–2464. https://www.atmos-chem-phys.org/acp/4/2449/.